\newcommand\sovast{\ref@jnl{Soviet~Ast.}} 

\RequirePackage[normalem]{ulem} 
\RequirePackage{color}\definecolor{RED}{rgb}{1,0,0}\definecolor{BLUE}{rgb}{0,0,1} 

\documentclass[12pt,usenatbib]{mn2e}
\usepackage{url,ulem,times,graphicx,amsmath,amsfonts,amssymb,color,epsfig,epstopdf}
\usepackage{graphicx}
\usepackage{epsf}
\usepackage{bm}
\usepackage{color}
\usepackage{rotating}
\usepackage[T1]{fontenc}
\usepackage{ae,aecompl}

\def\lsim{\mathrel{\lower0.6ex\hbox{$\buildrel {\textstyle <}
 \over {\scriptstyle \sim}$}}}
\def\gsim{\mathrel{\lower0.6ex\hbox{$\buildrel {\textstyle >}
 \over {\scriptstyle \sim}$}}}

\def\eone{${\bf e}_{1}$}
\def\etwo{${\bf e}_{2}$}
\def\ethree{${\bf e}_{3}$}

\begin{document}

\title[Universality of Subhalo Accretion ]{The universal nature of subhalo accretion}
\author[Libeskind et al]{Noam I Libeskind$^{1}$, Alexander Knebe$^{2}$, Yehuda Hoffman$^{3}$, Stefan Gottl\"{o}ber$^{1}$\\
$^1$Leibniz-Institute f\"ur Astrophysik Potsdam (AIP), An der Sternwarte 16, D-14482 Potsdam, Germany\\
  $^2$Grupo de Astrofisica, Departamento de Fisica Teorica, Modulo C-8, Universidad Aut\'onoma de Madrid, Cantoblanco E-280049, Spain\\
$^3$Racah Institute of Physics, Hebrew University, Jerusalem 91904, Israel\\
  }

\date{Accepted --- . Received ---; in original form ---}
\pagerange{\pageref{firstpage}--\pageref{lastpage}} \pubyear{2014}
\maketitle

 \begin{abstract}
We examine the angular infall pattern of subhaloes onto host haloes in the context of the large-scale structure. We find that this infall pattern is essentially driven by the shear tensor of the ambient velocity field. Dark matter subhaloes are found to be preferentially accreted along the principal axis of the shear tensor which corresponds to the direction of weakest collapse. We examine the dependence of this preferential infall on subhalo mass, host halo mass and redshift. Although strongest for the most massive hosts and the most massive subhaloes at high redshift, the preferential infall of subhaloes is effectively universal in the sense that its always aligned with the axis of weakest collapse of the velocity shear tensor. It is the same shear tensor that dictates the structure of the cosmic web and hence the shear field emerges as the key factor that governs the local anisotropic pattern of structure formation. Since the small (sub-Mpc) scale is strongly correlated with the mid-range ($\sim10$ Mpc) scale - a scale accessible by current surveys of peculiar velocities - it follows that findings presented here open a new window into the relation between the observed  large scale structure unveiled by current surveys of peculiar velocities and the preferential infall direction of the Local Group. This may shed light on the unexpected alignments of dwarf galaxies seen in the Local Group. \\

\end{abstract}

\section{Introduction}
\label{section:intro} 

\cite{1970A&A.....5...84Z} first introduced the concept of anisotropy in to the way we think about structure formation in cosmology. Yet, the dominant theoretical approach to structure formation, that still prevails to some extent today, has been based on spherical symmetry and the neglect of anisotropic dynamics. The most powerful quantitative measures of the growth of structure such as the two-point and higher order correlation functions \citep[e.g.][]{1980lssu.book.....P}, multiplicity, luminosity and mass functions \citep{2003ApJ...599...38B,2001MNRAS.321..372J} as well as analytical (and numerical) measures of merger-rates \citep{1993MNRAS.262..627L,1994MNRAS.271..676L} are all devoid of any reference to directions and anisotropy.  The spherical top-hat collapse model has been the corner stone, and reference point, to much of the analytical thinking on structure formation \citep{1974ApJ...187..425P,1999MNRAS.308..119S}.

Numerical $N$-body simulations have been, arguably, the main driving force of research on structure formation \citep[e.g.][among others]{2006Natur.440.1137S}. Even a causal visual inspection of cosmological simulations reveals the anisotropic nature of the growth of structure,  and evokes the "pancake" theory of Zeldovich and his co-workers \citep[e.g.][]{1982Natur.300..407Z,1980MNRAS.192..321D}. The analytical approach to galaxy formation, namely the cooling and fragmentation of gas in dark matter (DM) halos, relies even more heavily on the top-hat model and hence the assumption of spherical  spherical symmetry was integrated into the fabric of the theory of galaxy formation \citep{1977MNRAS.179..541R,1978MNRAS.183..341W}. These studies envisaged galaxy formation proceeding from the heating of gas accreted onto DM halos to virial temperatures and subsequently cooling and fragmenting to form stars. This picture has been recently challenged by \cite{2009Natur.457..451D} who argued that galaxy formation proceeds via cold narrow streams of gas that penetrate the shock-heated intra halo gas. \cite{2012MNRAS.422.1732D}, who studied the anisotropic infall pattern onto 350 DM halos of mass $\sim10^{12}M_{\odot}$ at redshift $z=2.5$, provided an extensive description of this infall pattern and attempted to relate it to the general properties of the cosmic web.

The notion of the cosmic web provides a very tempting framework for describing the anisotropic mass assembly of halos and galaxies. Subhaloes shape the the halo they inhabit \citep{2008ApJ...675..146F,2014arXiv1401.2060H}, and a number of  studies have shown how the orientation of galactic spin \citep{Aragon-Calvoetal2007,2009ApJ...706..747Z,2013ApJ...762...72T} or halo shape \citep{1982A&A...107..338B,2006MNRAS.370.1422A,2007MNRAS.375..184B,Libeskind2013a} is tied to large scale structure. \cite{Libeskind2013b} found that halo spin aligns itself with the cosmic vortical field, while a number of related studies found weaker alignments with the ``cosmic web''. These numerical approaches have been complimented by a number of recent observational studies that have found similar trends in redshift surveys \citep{2013MNRAS.428.1827T,2013ApJ...775L..42T,2014MNRAS.437L..11T,2014arXiv1403.5563G}. Yet, the notion of a web that categorizes the structure into four distinct elements of voids, sheets, filaments and knots is somewhat ill defined and arbitrary. The fuzziness of the web classification and its arbitrary nature hamper the attempt to understand and relate the anisotropic nature of the mass assembly of halos and galaxies to the inherent anisotropic nature of the cosmic web. This calls for a new approach which relies on a robust characterization of the large scale structure free of arbitrary fine tuning and thresholds adjustments. One that can be easily defined in a scale free way across different mass scales and different redshifts.    

In this work the LSS is defined using the eigenvectors of the velocity-shear field. Such a definition is ``democratic'' in the sense that each point in space has an equally well-defined LSS irrespective of other environmental factors such as density. Using this definition we show that the accretion of subhaloes onto host haloes is universally reflective of the shear field.

\section{Method}

In order to examine the anisotropy - if one exists - of the angular infall pattern of subhaloes crossing the virial sphere of their host haloes, we use a DM-only $N$-body simulation of 1024$^3$ particles in a 64$h^{-1}$ Mpc box. Such a simulation achieves a mass resolution of $1.89\times10^{7}h^{-1}$M$_{\odot}$ per particle and a spatial softening length of 1 $h^{-1}$kpc. A standard WMAP5 (Komatsu et al 2009) $\Lambda$CDM cosmology is assumed: $\Omega_{\Lambda}=0.72,~\Omega_{\rm m}=0.28,~\sigma_{8}=0.817$ and $H_{0}=70$km/s/Mpc. The publicly available {\sc Gadget2} (Springel et al 2005) code is used and 190 snapshots (equally spaced in expansion factor) are stored from $z=20$ to $z=0$. The same simulation was used in Libeskind et al (2012). 

The velocity shear field is defined by the symmetric tensor: 
$$\Sigma_{ij}=-\frac{1}{2H(z)}\bigg(\frac{\partial v_{i}}{\partial r_{j}}+\frac{\partial v_{j}}{\partial r_{i}}\bigg)
$$where $i,j=x,y,z$. The $H(z)$ normalization is used to make the tensor dimensionless and the minus sign is introduced to make positive eigenvalues correspond to a converging flow. As dictated by convention, the eigenvalues are sorted in increasing order ($\lambda_{1} > \lambda_{2} > \lambda_{3}$), and the associated eigenvectors are termed \eone, \etwo, and \ethree. Note that \cite{Hahnetal2007a} and \cite{Forero-Romero2009} suggested that the number of eigenvalues above a specified threshold may be used to identify the constituent elements of the cosmic web namely, voids, sheets, filaments and knots.

In order to compute the shear tensor at each point in the simulation, the velocity field is gridded according to a  Clouds-In-Cell (CIC) scheme. This is then smoothed with a gaussian kernel in Fourier space. The shear is then computed by means of an FFT. Note that other methods for extracting the velocity field exist in the literature, e.g. by Delaunay tessellation \citep[i.e.][among others]{2013MNRAS.429.1286C}. The size of the CIC used here is 256$^3$, chosen such that every mesh cell contains at least one particle at $z=0$ .The width of the gaussian smoothing we apply is adaptive and depends on the mass of the halo we wish to examine (see below).

Host and sub-haloes are identified by means of the publicly available halo finder {\sc AHF} \citep{Knollman09}. At each redshift $z<5$ host haloes are divded into five mass bins from $10^{9}$ to $10^{14}h^{-1}M_{\odot}$, each a decade wide. The median virial radius for each mass bin is then computed. For each halo we choose to employ the shear field smoothed 4, 8, and 16 times the median virial radius of the mass bin the halo is in. In this way we ensure that the eigen-frame employed is always adapted to the host halo such that massive haloes and small haloes are treated equally.

Accretion events are found by identifying which subhaloes at a given snapshot $z_{1}$ are identified as ``field'' haloes at the previous snapshot $z_2$ (where $z_{1}<z_{2}$), by building a Merger-tree. The accreted position is assumed to be the midpoint between the subhaloes position (with respect to the host) at $z_{1}$ and $z_{2}$. Every time an accretion event is found, its position with respect to the eigenvectors of the shear tensor is computed and recorded.

A small fraction of the accreted sub-halos enter, exit and then re-enter the halo more than once. This multiple-entry phenomenon jeopardizes the counting statistics of in falling substructures and thus needs to be properly subtracted. This is done by tracking all accreted subhaloes back in time through their Merger-tree and checking if they were ever identified as subhaloes at any previous redshifts. For the small fraction of subhaloes where this is phenomena occurs, only the first entry is considered for the statistical analysis presented here.

Much work has been done on this phenomenon \citep[e.g.][]{2004A&A...414..445M,2005MNRAS.356.1327G,2008MNRAS.385.1859W,2012ApJ...751...17S}. Most studies find that such multiple entry events are most common for high mass haloes and rarer for low mass ones. The number of ``re-enterers'' found in our simulation (although mass, resolution and redshift dependent) is fully consistent with these previous studies, and are omitted from this analysis entirely.

Following  \cite{2008MNRAS.385..545K}, host halo mass is scaled by $M_{\star}(z)$, the mass of a typically collapsing object at a given redshift, namely $\widetilde{M} = M_{\rm halo} / M_{\star}$. $M_{\star}(z)$ is defined by requiring that the variance $\sigma^{2}$, of the linear over-density field within a sphere of radius $R(z)=(3M_{\star}(z)/4\pi\rho_{\rm crit})^{1/3}$, should equal to $\delta{_{c}^{2}}$, the square of the critical density threshold for spherical collapse  \citep[e.g. see][]{1974ApJ...187..425P,1997ApJ...490..493N,2008MNRAS.385..545K}. $M_{\star}(z)$ is calculated using the cosmological parameters adopted here: at the present epoch $M_{\star}=3.6\times10^{12}h^{-1}M_{\odot}$. At $z=5$, $M_{\star}(z=5)\approx10^{8}h^{-1}M_{\odot}$.

\begin{figure*}
 \includegraphics[width=40pc]{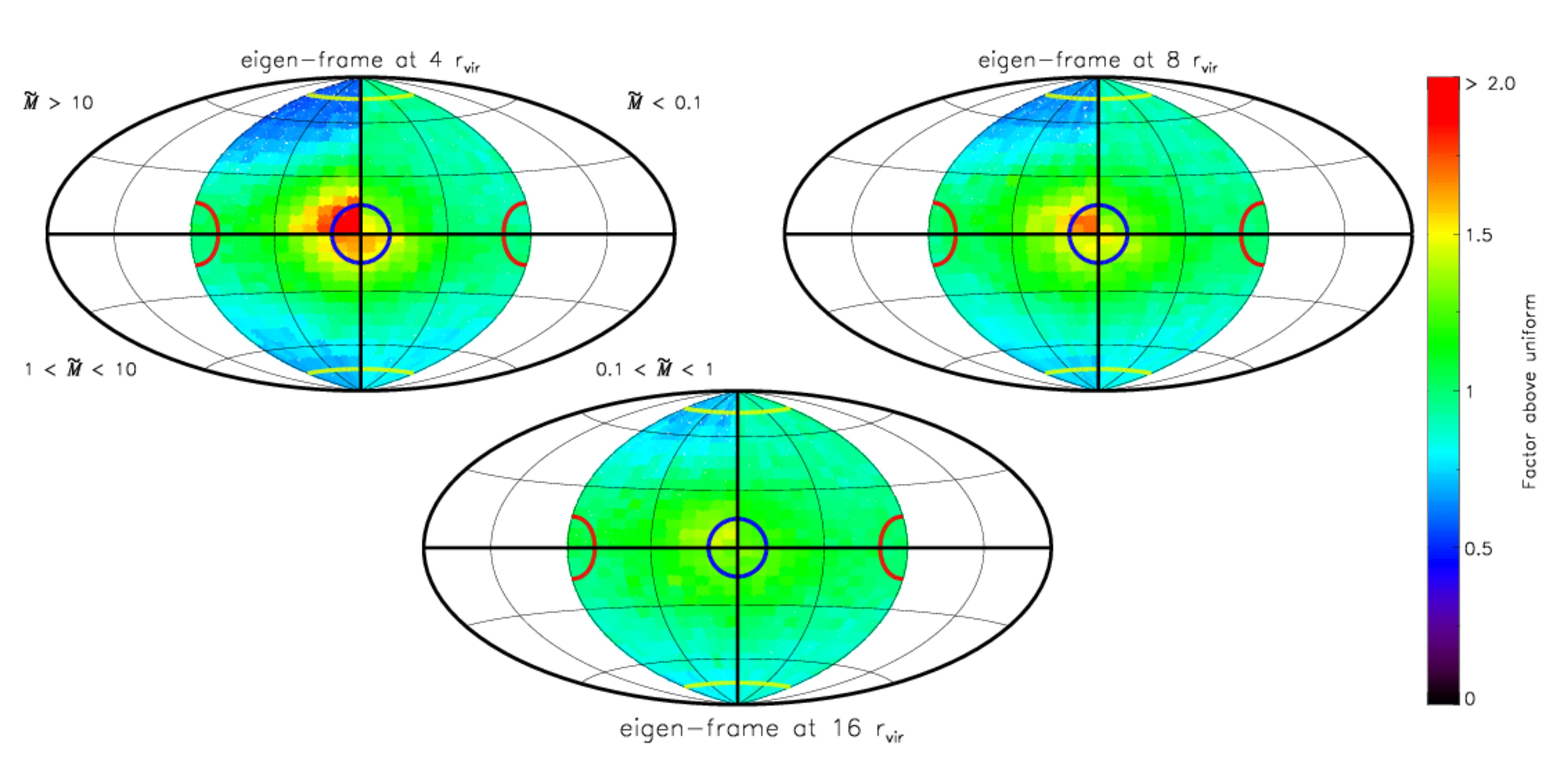}
 \caption{The location of subhalo entry points is shown in an Aitoff projection of the virial sphere. The density of subhalo entry points is shown for eigen-frames smoothed on 4 (upper left), 8 (upper right) and 16 (bottom) virial radii. Starting from ``noon'' and going counter-clockwise, we show these entry points for accretion events occurring on to host haloes in four different mass ranges $\widetilde{M} <0.1;0.1<\widetilde{M} <1;1<\widetilde{M} <10$ and $\widetilde{M} >10$ and at all redshifts below z $\sim$ 5. $\widetilde{M}$ is a measure of the halo mass in units of the mass of a collapsing object at each redshift. The density of entry points is normalized to that expected from a uniform distribution, and contoured accordingly. The ``north'' and ``south'' pole correspond to \eone; the two mid points on the horizontal axis at $\pm180^{\circ}$ to correspond to \etwo, while the midpoint corresponds to \eone. The yellow, red and blue circles define areas within 15 degrees of the eigen-frame axes, \eone, \etwo, and \ethree, respectively. \vspace {1cm}\newline The distribution of entry  points is never consistent with uniform. Instead it universally (irrespective of host halo mass, scale on which the shear is computed or redshift) peaks close to \ethree: on large scales the shear tensor dictates the shape of cosmic web and on small scales it determines the infall pattern of satellites}
 \label{figure:maps}
 \end{figure*}

\section{Results}
The eigenvectors of the shear tensor, evaluated at the position of each host halo, provides the principal orthonormal vectors within which the anisotropy of mass aggregation onto halos can be naturally examined. Because the eigenvectors are orthonormal, they define an ``eigen-frame''. Each halo has its own eigen-frame, defined by the ambient shear field.

Given that the eigenvectors are non directional lines, this corresponds to a single octant of the 3D cartesian coordinate system. The location of where subhaloes cross the halo virial radius (``entry points'') is plotted in this eigen-frame. In Figs.~1 and 2 we stack the accretion events onto all host haloes at all redshifts. Fig.~1 shows the entry points in an Aitoff projection for all accretion events while Fig.~2 shows these for mergers where the subhalo mass is greater than 10\% of the host.

\begin{figure*}
 \includegraphics[width=40pc]{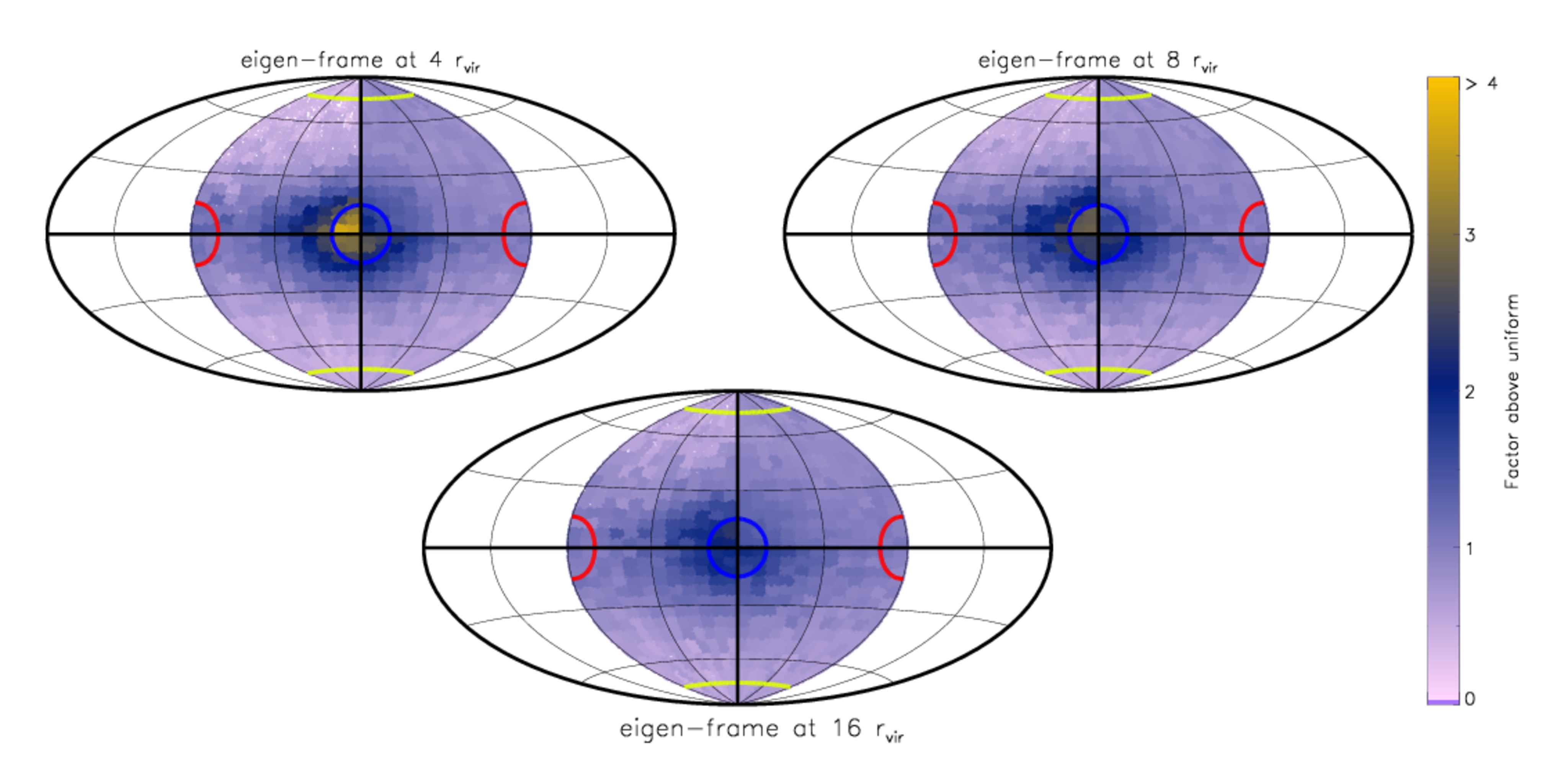}
 \caption{Same as Fig.~1 but only considering subhaloes whose mass is greater than 10\% of their host.}
 \label{figure:maps}
 \end{figure*}
In each of these figures we show the entry points in the eigenframe defined by the shear computed with three different smoothings that correspond to 4 (upper left), 8 (upper right) and 16 (bottom) virial radii. Host haloes are divided into four mass bins according to $\widetilde{M}$. Starting at ``noon'' and going clockwise, these are where $\widetilde{M} <0.1$; $0.1<\widetilde{M} <1$; $1<\widetilde{M} <10$; and $ 10 <\widetilde{M} $. In order to quantify the statistical significance of any anisotropy in the angular entry-point distribution, we divide the number of entry points in a given area on the virial sphere by that expected from a uniform distribution. 

Owing to the high resolution of our simulation we obtained 883,245 accretion events since $z=5$. The four mass bins have been chosen to have roughly equal number of accretion events. Therefore, at each point on the virial sphere, the variance due to Poisson statistics of a uniform distribution of the same number of points, is small.

\begin{figure*}
 \includegraphics[width=42pc]{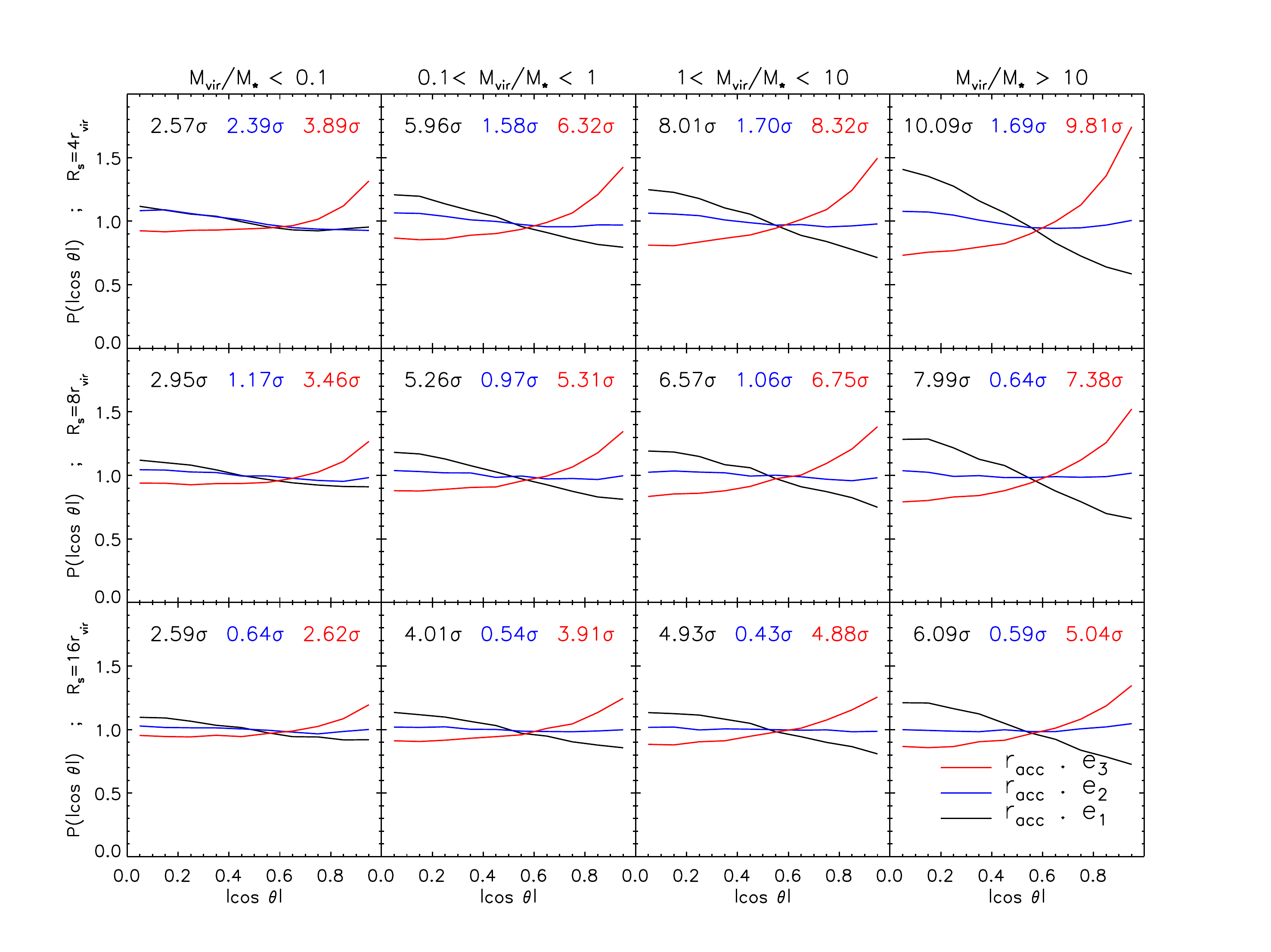}
 \caption{The anisotropic accretion shown in Fig.~1 is quantified by means of a probability distribution, $P(|\cos\theta|)$ of the cosine of the angle made between a subhalo's entry point (${\rm \bf r} _{\rm acc}$) and the eigenvectors \eone~(black), \etwo~(blue) and \ethree~(red). The top, middle and bottom rows show the probability distribution when the shear has been smoothed on 4, 8 and 16$r_{\rm vir}$. The probability distributions are split according to value of $\widetilde{M}$, denoted on top of each column. The statistical significance of each probability distribution is characterized by the average offset between it and a random distribution in units of the Poisson error and is indicated by the corresponding colored number in each panel. Distributions that are consistent with random have values $ < 1\sigma$.}
\label{fig:Prob-dist}
 \end{figure*}

Since the three eigenvectors (\eone, \etwo, and \ethree) of the shear tensor define the coordinate system used to construct these Aitoff projections, we plot yellow, red and blue circles demarcating $15^{\circ}$ about each of axes, respectively in order to highlight the universal nature of the anisotropic accretion.

Fig.~1 and 2 show that there is a strong tendency for the accretion to occur along \ethree. Regardless of the host halo mass, the merger ratio or the smoothing used, there is a statistically significant tendency for subhaloes to be accreted closer to \ethree~than to either other of the eigenvectors. Recall that \ethree~corresponds to the direction of slowest collapse. This is the main result of this paper: {\it subhaloes are preferentially accreted along the direction that corresponds to slowest collapse.} Note that this effect is greatest for the most massive host haloes and becomes progressively weaker as halo mass decreases. 
Also, as the gaussian smoothing kernel is increased the effect also weakens. This is expected: large smoothing kernels effectively homogenize the LSS, randomizing the principal direction of the shear tensor. Finally the tendency to be accreted along \ethree~is largest for ``massive'' subhaloes that are greater than 10\% of their hosts. In the ``best'' case, where the smoothing is confined to 4$r_{\rm vir}$, where only the most massive host haloes and the greatest merger events are considered, the mergers are more than 4 times as likely to come along \ethree, than expected from a uniform distribution.

By stacking our results in the manner shown, we have explicitly omitted any dependence of subhalo accretion on redshift or absolute halo mass. Below we examine how the funneling of accretion events changes with redshift and host halo mass. 

\begin{figure*}
 \includegraphics[width=40pc]{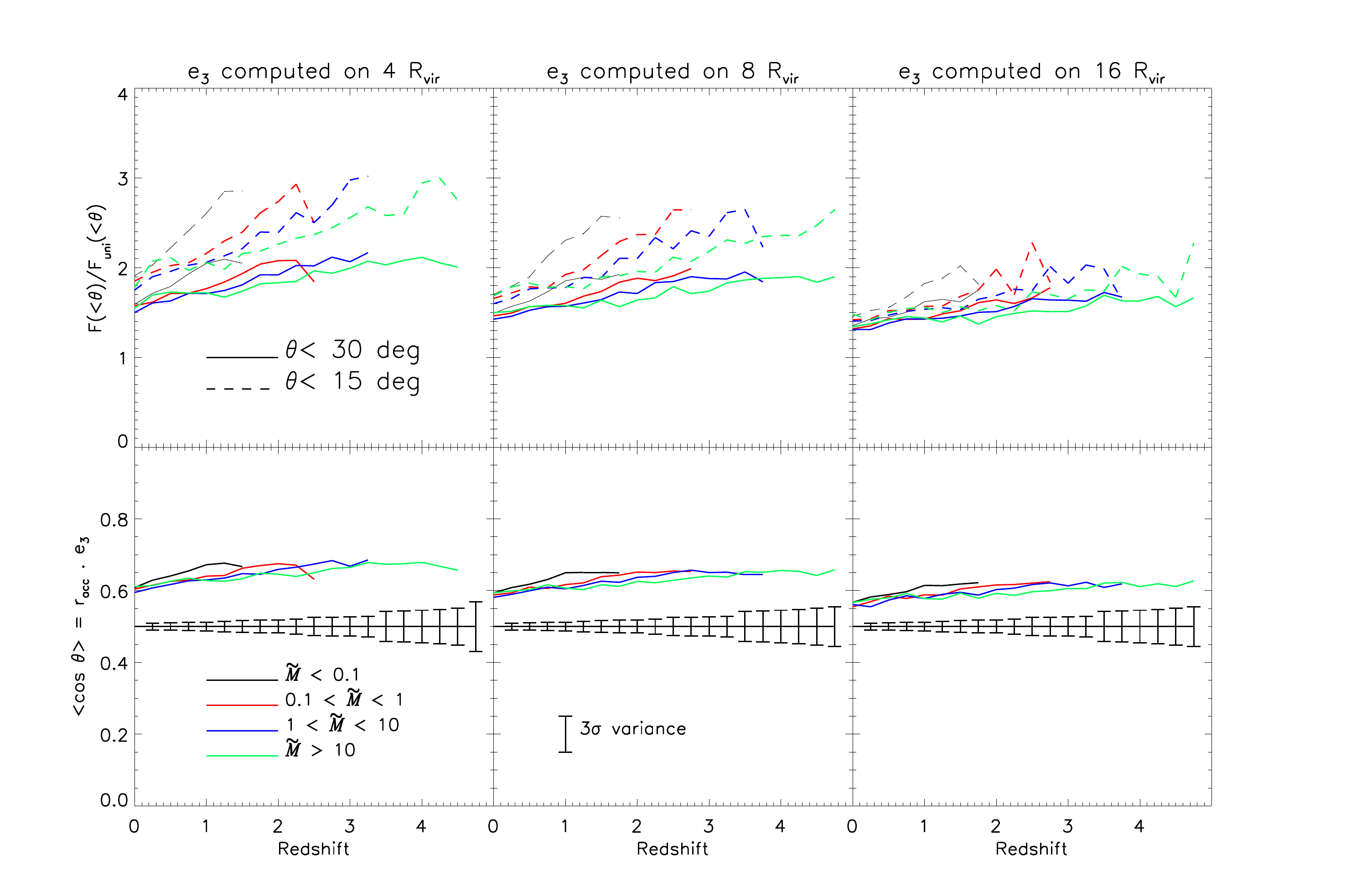}
 \caption{The beaming of subhalo accretion narrows with increasing redshift. In the upper panels we show the fraction of subhaloes that are accreted within 15 and 30 degrees (dashed and solid lines, respectively) of \ethree~normalized by the fraction expected from a spherically uniform distribution. Each of the the four host halo mass bins are colored according to the legend. In the bottom panels we show the median angle formed between the position vector of a given subhalo at the moment of accretion (${\rm \bf r}_{\rm acc}$)  and \ethree. The error bars in these plots represent the 3$\sigma$ variance of the median angle found, given 10,000 same sized, uniform distributions at each redshift (namely, when 10,000 medians are computed from 10,000 uniform distributions the error bars represent the 0.15 and 99.85 percentiles in the distribution of medians). The left, middle and right columns show how these quantitates vary as the shear tensor (and its eigenvectors) are computed with increasing scale namely 4, 8 and 16$r_{\rm vir}$, respectively. Note that only the greatest mass bin extends to $z=5$ since less massive host haloes at high redshift are (relatively) small and thus nearly all accretion is poorly resolved.}
  \label{fig:zdepen}
 \end{figure*}

In Fig.~\ref{fig:Prob-dist} we show the probability distribution of the cosine of the angle formed between the subhalo entry point (${\bf r}_{\rm acc}$) and the eigenvectors of the shear \eone~(black), \etwo~(blue) and \ethree ~(red), namely $\cos\theta={\bf r}_{\rm acc}\cdot~{\bf e}_{i}$,
where $i=1,2,3$. In what follows only this angle is considered. The probability distributions in Fig.~\ref{fig:Prob-dist} are valid for all redshifts but are split by mass (first column: $\widetilde{M} <0.1$; second column: $0.1<\widetilde{M} <1$; third column: $1<\widetilde{M} <10$ and fourth column $\widetilde{M} >10$). Additionally the eigenvectors are computed on three scales corresponding to smoothing of 4 (top), 8 (middle) and 16 (bottom row) times the halo's virial radius. Uniform distributions would be represented by a solid flat line at unity.  

The strength (or weakness) of the alignment can be characterized by $\sigma$, the average offset between a given probability distribution and a random one, calculated in units of the Poisson error. In practice, the average difference between the number of entry points found in a given bin and the number expected
from a uniform distribution is calculated in terms of the Poisson error of a random distribution of the same size. If $\sigma$ is less than unity, then (on average) the measured alignment lies within the Poisson error of a uniform distribution. High values of $\sigma$ indicates a strong deviation from uniformity while a statistically weak signal corresponds to low values of $\sigma$. In each panel we indicate the strength (or weakness) of the alignment signal.

The same trends seen in Fig. 1 and 2 are seen here too: the beaming of subhaloes is strongest when the velocity shear tensor is computed on the smallest scales, and when accretion onto the largest hosts is considered. Note that at all smoothing lengths and at all masses, subhaloes are statistically aligned with \ethree, and away from \eone. For some host halo mass bins and smoothing lengths, the alignment of infall points with \etwo~is not statically signifiant. 

The funneling or beaming of substructures is mildly redshift dependent as shown in Fig.~\ref{fig:zdepen}. In the top panels we show the fraction of subhaloes that are accreted within 15 and 30 degrees (dashed and solid lines, respectively) above what is expected from a uniform distribution, as a function of redshift for the four mass bins of $\widetilde{M}$. As inferred from the previous plots, the beaming effect lessens as the shear tensor is computed on larger and larger scales. Accretion at high redshift is more aligned with \ethree, than accretion at low redshift. This is true for all smoothing scales and for all mass bins. Nevertheless, despite the widening of the subhalo ``funnel'' at low redshift, the likelihood that a subhalo is accreted close to \ethree~ is still well above that expected from a uniform distribution. This is evident also when examining the median (cosine) angle formed between \ethree~and $\bf{r}_{\rm acc}$, the position vector of each subhalo at the moment it crosses the virial radius (shown in the lower row of Fig.~\ref{fig:zdepen}). In these figures it is quite clear that the median angle is not close to 0.5, that which would be expected from a uniform distribution and thus represents a statistically significant alignment. 

We can quantify how likely it is to obtain such a setup due to random accretion, by constructing 10,000 uniform distributions between $[0,1]$, at each redshift and of identical size to the number of accretion events at that redshift. We then sort these 10,000 medians and examine the 3$\sigma$ spread of these values (namely the 0.15 and 99.85 percentile of the distribution of medians). This is shown as the error bars about 0.5 (the ``median'' median, or 50th percentile of the distribution). Note that these error bars increase in size at higher redshift since the number of accretion events decreases due to the simulation's resolution. At higher redshift there are are fewer resolved haloes (above a given mass) and thus the number of accretion events necessarily decreases. Regardless, even at high redshift when the only resolvable accretion events are on to the most massive host haloes, they still occur well aligned with \ethree.

Although not shown here we have examined the dependence of the alignment on cosmic web environment. Namely, following Hahn et al (2007), a halo may be classified as existing in a knot, filament, sheet or void by counting how many eigenvalues are positive. No dependence on web environment is found according to this scheme.

\section{Summary and Discussion}

It has long been realized that DM halos grow in an anisotropic fashion. This has often been claimed to be related to the cosmic web which constitutes the scaffolding for the building of halos and the galaxies within them. The accretion of matter onto halos generally proceeds in two modes: the accretion of clumps and the  smooth accretion of diffuse material. In this paper, we have focused on the mergers of small halos with massive ones and analyzed their anisotropic infall pattern with respect to the eigen-frame of each individual host halo. The eigen-frame is defined by the three eigenvectors of the velocity shear tensor evaluated at the position of each halo. Our main finding is that, across a range of halo masses and redshifts, the infall direction of subhaloes is preferentially confined to the plane orthogonal to \eone~(the direction of fastest collapse), within which it is aligned with \ethree~(the axis of slowest collapse). 

Some of the main characteristics of the infall of subhalos onto more massive halos are listed here:
\begin{itemize}
\item Subhaloes tend to be accreted onto hosts from a specific direction with respect to the large scale structure.
\item In the case of filaments, the \ethree\ direction 
coincides with the ``spine'' of the filament \citep{2014MNRAS.437L..11T}. Hence, the well known phenomenon of halos being fed by substructures funneled  by filaments is recovered here. 
\item The strength of the beaming effect depends somewhat on the length scale used to compute the velocity-shear eigenframe: the smaller the scale, the stronger the beaming. 
\item More massive subhaloes are more anisotropically accreted onto host haloes than smaller subhaloes.
\item Similarly, more massive host haloes accrete subhaloes more anisotropically than smaller ones.
\item Accretion at high redshift is more anisotropic than accretion at low redshift, for all masses of hosts and subhaloes.
\end{itemize}

A somewhat naive reasoning might suggest that halos should be nourished along \eone, the axis of the fastest collapse. However, this reasoning is flawed: halos grows by accreting material from their surrounding, and this occurs most rapidly in the direction of \eone. It follows that at the time a given halo is inspected, the mere existence of that halo implies that much of the surrounding material has already been consumed by the halo along \eone. This leaves the material along \ethree\ as the main supply of fresh material that feeds the halo.

The beaming of subhalo accretion onto halos in a given bin of $\widetilde{M}$ (namely $M_{\rm vir}$ scaled by the redshift dependent $M_{\star}(z)$), narrows with increasing redshift (Fig. 4). In a scale-free universe, i.e. an Einstein-de Sitter cosmology with a power law power spectrum, the angular dependence of the accretion is expected to be completely epoch independent. This does not hold for the $\Lambda$CDM cosmology assumed here. As the universe evolves it becomes more dominated by the $\Lambda$ term and consequently the role of gravity (via subhalo dynamics) diminishes with time. This is manifested in the accretion and merger rate: accretion onto halos decreases and the funneling of matter along \ethree\ gets weaker.

Arguably, the most important ramification of this paper is that anisotropic nature of the mass growth of halos is dictated by the velocity shear tensor and not by cosmic web (Hoffman et al 2012). That is to say, the anisotropic nature of subhalo accretion {\it does not} depend on the magnitude of the shear tensor's eigenvalues, nor does it depend on the ``web environment''. The beaming of subhaloes along ${\bf e}_{3}$ occurs equally in knots, filaments, sheets and voids. Rather, the shear tensor is the one that characterizes, shapes and dictates the directions of the cosmic web. This provides further support to earlier claims regarding the dominance of the shear tensor in shaping the large scale structure \cite{2012MNRAS.421L.137L,Libeskind2013b}

\cite{2014MNRAS.441.1974L} have recently shown that the principal directions of the shear tensor remain coherent over a wide range of redshifts and spatial scales. This opens  interesting possibilities for relating the observed large scale velocity field with the properties of halos, and hence of galaxies and groups of galaxies.  The work presented here, combined with observations of the local velocity field, will allow us to thus identify the direction along which most accretion onto the Local Group occurred. Such findings can have important implications on the peculiar geometric set up of dwarf galaxies in the local group.

\section*{Acknowledgments}
NIL is supported by the Deutsche Forschungs Gemeinschaft. AK is supported by the {\it Ministerio de Econom\'ia y Competitividad} (MINECO) in Spain through grant AYA2012-31101 as well as the Consolider-Ingenio 2010 Programme of the {\it Spanish Ministerio de Ciencia e Innovaci\'on} (MICINN) under grant MultiDark CSD2009-00064. He also acknowledges support from the {\it Australian Research Council} (ARC) grants DP130100117 and DP140100198. YH has been partially supported by the Israel Science Foundation (1013/12). SG and YH have been partially supported by the Deutsche Forschungsgemeinschaft under the grant $\rm{GO}563/21-1$. The simulations have been performed at the Leibniz Rechenzentrum (LRZ) in Munich. This paper was supported (in part) by the National Science Foundation under Grant No. PHY-1066293 and the hospitality of the Aspen Center for Physics.

\bibliography{./ref}
\end{document}